\documentclass[aps,twocolumn,doi=false,pra,amsmath]{revtex4-1}
\usepackage{graphics}
\usepackage{natbib}
\usepackage{graphicx}
\usepackage{epsfig}
\usepackage{amsmath}
\usepackage{bm}
\usepackage{color}

\begin{document}
\title{Spin injection and detection up to room temperature in Heusler~alloy/\textit{n}-GaAs spin valves}

\author{T. A. Peterson$^1$, S. J. Patel$^2$, C. C. Geppert$^1$, K. D. Christie$^1$, A. Rath$^4$, D. Pennachio$^2$, M. E. Flatt{\'{e}}$^3$, P. M. Voyles$^4$, C. J. Palmstr{\o}m$^2$, and P. A. Crowell$^1$}
\date{\today}
\affiliation{$^1$School of Physics and Astronomy, University of Minnesota, Minneapolis, Minnesota 55455\\
$^2$Departments of Electrical \& Computer Engineering and
Materials, University of California, Santa Barbara, California
93106 \\ $^3$Department of Physics and Astronomy, University of
Iowa, Iowa City, Iowa 52242 \\ $^4$Department of Materials Science
and Engineering, University of Wisconsin-Madison, Madison, WI
53706 }
\begin{abstract}
We have measured the spin injection efficiency and spin lifetime
in Co$_2$FeSi/\textit{n}-GaAs lateral nonlocal spin valves from 20
to 300 K. We observe large ($\sim$40 $\mu$V) spin valve signals at
room temperature and injector currents of $10^3$~A/cm$^2$,
facilitated by fabricating spin valve separations smaller than the
1 $\mu$m spin diffusion length and applying a forward bias to the
detector contact. The spin transport parameters are measured by
comparing the injector-detector contact separation dependence of
the spin valve signal with a numerical model accounting for spin
drift and diffusion. The apparent suppression of the spin
injection efficiency at the lowest temperatures reflects a
breakdown of the ordinary drift-diffusion model in the regime of
large spin accumulation. A theoretical calculation of the
D'yakonov-Perel spin lifetime agrees well with the measured
\textit{n}-GaAs spin lifetime over the entire temperature range.

\end{abstract}
\maketitle
\section{Introduction}
All-electrical spin transport has been demonstrated in III-V
semiconductors \cite{Lou2006a, Ciorga2009, Salis2010,Saito2013},
group IV semiconductors \cite{Appelbaum2007}, and in 2D materials
such as graphene \cite{Tombros2007, Han2010}. One of the most
mature systems studied in the field of semiconductor spintronics
is the ferromagnet (FM)/\textit{n}-GaAs lateral spin valve (SV)
structure \cite{Lou2006a, Ciorga2009, Salis2010}. GaAs-based
devices have served as a testbed for several seminal semiconductor
(SC) spin transport measurements, such as the Hanle effect
\cite{Lou2006a, Lou2006b}, the spin Hall and inverse spin Hall
effects \cite{Kato2004a, Werake2011, Garlid2010a}, and nuclear
hyperfine effects \cite{Lou2006b, Chan2009, Awo-Affouda2009,
Salis2009}. The Dresselhaus spin-orbit interaction (SOI)
\cite{dresselhaus1955} originating from the non-centrosymmetric
lattice of III-V SCs makes them attractive candidates for
modulation of spin transport using the SOI \cite{Datta1990}. At
the same time, however, the Dresselhaus SOI present in III-V SCs
leads to efficient spin relaxation in the diffusive transport
regime.

Electron spin relaxation in \textit{n}-GaAs at doping levels near
the metal-insulator transition is governed by the D'yakonov-Perel
(DP) mechanism \cite{DP1971,Dzhioev2002}. The DP spin relaxation
rate in III-V semiconductors has a characteristic $\tau_s^{-1}
\propto \epsilon^3$ behavior \cite{DP1971,MeierZak1984}, where
$\epsilon$ is the carrier energy. The spin lifetime $\tau_s$ is
the inverse of the spin relaxation rate. At temperatures for which
the carriers are nondegenerate ($\epsilon \sim k_b T$), the spin
lifetime falls sharply as $\tau_s \propto T^{-3}$
\cite{Kikkawa1998}. Short spin lifetimes ($\sim 10 - 100$~ps) have
therefore challenged \textit{n}-GaAs SV room temperature
performance \cite{Saito2013}, as the short spin lifetime limits
the steady-state spin accumulation.

In this article we demonstrate electrical detection of nonlocal
spin accumulation in Heusler alloy FM/\textit{n}-GaAs lateral spin
valve devices up to room temperature. Clear nonlocal SV signals
are measured by fabricating devices with injector-detector contact
separations of less than a spin diffusion length and applying a
forward bias voltage to the detector contact. We use the
injector-detector contact separation dependence of the SV signal
to extract the \textit{n}-GaAs spin lifetime and FM/SC interface
spin injection efficiency from 20 K up to room temperature. These
data allow for a comprehensive and quantitative evaluation of the
temperature-dependent performance of FM/\textit{n}-GaAs lateral SV
devices.  We find that the spin lifetime in the \textit{n}-GaAs
channel is in quantitative agreement with a theoretical
calculation of the DP spin lifetime over the entire temperature
range. At low temperatures, we achieve a spin accumulation that is
a significant fraction of the carrier density in the channel. This
is accompanied by an apparent downturn in the injection efficiency
which we believe is due to breakdown of the ordinary
drift-diffusion model in the regime of large spin-dependent
electrochemical potential splitting.
\section{Methods}
\subsection{Structure growth and device fabrication}
\begin{figure}
\includegraphics{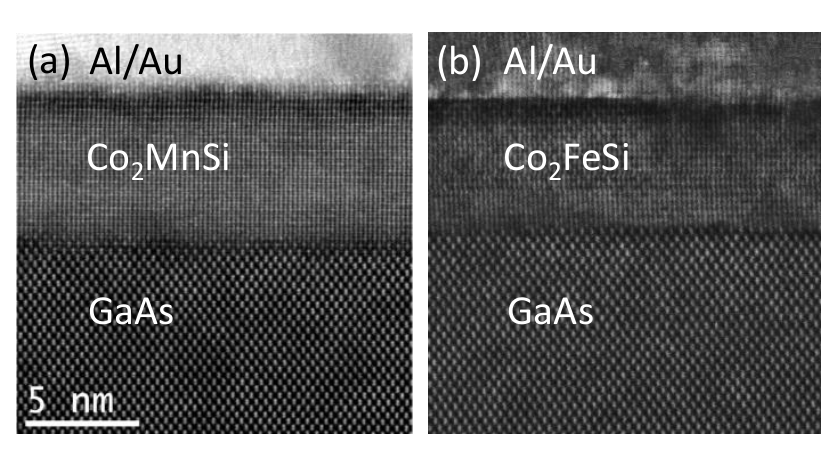}
\caption{Cross-sectional HAADF-STEM images of (a) the
Co$_2$MnSi/GaAs interface and (b) the Co$_2$FeSi/GaAs interface.
Images (a) and (b) were taken on the same heterostructures used
for the Co$_2$MnSi and Co$_2$FeSi spin valves measurements
presented in this paper. A 5~nm scale bar is indicated in the
lower left of (a). } \label{fig:STEM}
\end{figure}

The devices used in this study were fabricated from
heterostructures grown by molecular-beam epitaxy (MBE). A 2.5
$\mu$m Si-doped ($n=3\times10^{16}$ cm$^{-3}$) GaAs epilayer was
grown following a 500~nm undoped GaAs buffer layer grown on a
semi-insulating (001) GaAs substrate. To thin the naturally
occurring Schottky depletion layer and provide a tunnel barrier
for efficient spin injection \cite{Rashba2000, Fert2001,
Hanbicki2002}, the doping level was increased at the FM/SC
interface. A 15 nm transitional doping layer was grown
($n=3\times10^{16}$ cm$^{-3}\rightarrow n^+=5\times10^{18}$
cm$^{-3}$) on top of the \textit{n}-GaAs epilayer, followed by an
18 nm thick heavily doped ($n^+=5\times10^{18}$ cm$^{-3}$) layer.
Following the GaAs MBE growth, the sample was cooled to
$<$~400$^\circ$~C under As$_4$-flux at which point the As$_4$-flux
was turned off. This resulted in a highly ordered GaAs(001)c(4x4)
As-rich surface reconstruction as confirmed by reflection
high-energy electron diffraction (RHEED) and \textit{in situ}
scanning tunneling microscopy (STM). For the 5~nm thick epitaxial
Heusler film growth, the samples were transfered to a separate
growth chamber while maintaining ultra-high vacuum (UHV). The
Heusler film growth was performed at 270$^\circ$~C with
codeposition from individual elemental sources. The Heusler
compounds grow with a cube-on-cube orientation with
Heusler(001)$<$110$>\vert\vert$~GaAs(001)$<$110$>$\cite{Hashimoto2005,hirohata2005}.
During Heusler growth RHEED was used to confirm layer-by-layer
growth of a single crystal film. Cross-sectional high-angle
annular dark field scanning transmission electron microscopy
(HAADF-STEM) was performed, and example images of the interfaces
are shown in Fig.~\ref{fig:STEM}. These images confirm the samples
are single crystals with mixed L2$_1$ and B2 phases in both
Co$_2$MnSi (Fig.~\ref{fig:STEM}(a)) and Co$_2$FeSi
(Fig.~\ref{fig:STEM}(b)) films, and a degree of intermixing at the
GaAs/Heusler interface of no more than 4-6 atomic layers. The
GaAs(001)/Heusler interface resulted in a uniaxial magnetic
anisotropy yielding an easy axis along the GaAs [110] direction
\cite{Hashimoto2005,Liu2014,Liu2016} for both the Co$_2$FeSi and
Co$_2$MnSi films.

\begin{figure}
\includegraphics{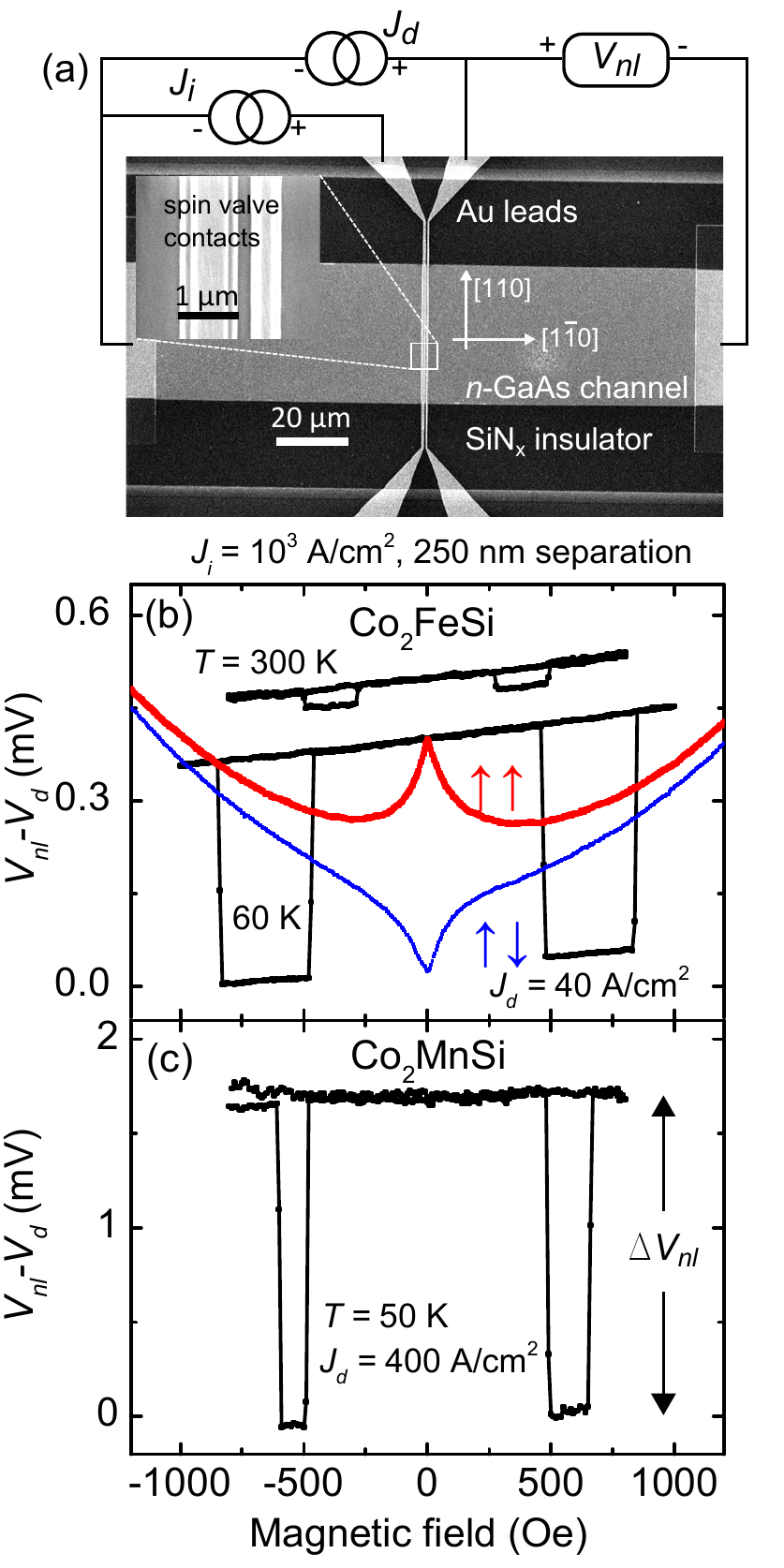}
\caption{(Color online) (a) Scanning electron micrograph of a
lateral SV device, with a schematic diagram of the measurement.
The inset is a magnified image of the injector (left contact) and
detector (right contact), in the device pictured with an
edge-to-edge separation of 250 nm. (b-c) Example BDSV field sweeps
for devices with Co$_2$FeSi contacts (b) and Co$_2$MnSi contacts
(c). The temperature and bias conditions are indicated on the
figure. $\Delta V_{nl}$ is the magnitude of the
parallel-antiparallel difference as indicated in (c). At the bias
conditions indicated in (b) $V_{d} = 0.44$ V at 60 K and $V_{d} =
0.30$ V at 300 K. In (c) $V_{d} = 0.72$ V at 50 K for the bias
conditions indicated. After subtracting $V_{d}$, the 60 K and 300
K data in (b) are offset for clarity. In (b), the dc NLH
measurement is shown at 60 K, for both parallel (red) and
antiparallel (blue) magnetization configurations.}
\label{fig:diagram}
\end{figure}

The heterostructures were patterned into lateral spin valve
devices using a top-down fabrication process. A combination of
electron-beam lithography and photolithography was used, with
Ar$^+$ ion milling to define the ferromagnetic contacts and wet
etching to define the \textit{n}-GaAs channel. A silicon nitride
insulating layer was deposited by plasma-enhanced chemical vapor
deposition (PECVD) and patterned by lift-off to electrically
isolate the evaporated Ti/Au vias and bonding pads from the
substrate and \textit{n}-GaAs channel sidewalls. A micrograph of a
SV device is shown in Fig.~\ref{fig:diagram}(a). The channel width
in the GaAs [110] direction is 80 $\mu$m, the SV contact length is
50 $\mu$m, the injector width is 1 $\mu$m, and the detector width
is 0.5 $\mu$m. The large aspect ratio of the SV contacts along the
magnetic easy axis was chosen in order to minimize fringe magnetic
fields as well as to define a two-dimensional geometry conducive
to modeling (channel width $\gg$ spin diffusion length). The
large-area remote contacts share the same composition as the SV
contacts. The remote contacts, however, have no impact on the SV
measurement, because they are placed many spin diffusion lengths
away from the SV contacts. Multiple SV devices were fabricated on
the same chip by wet etching through the 2.5 $\mu$m
\textit{n}-GaAs to isolate the devices electrically. SV devices on
the same chip were patterned with injector-detector edge-to-edge
separations ranging from 250 nm to 5 $\mu$m.

\subsection{Charge transport}
\begin{figure}
\includegraphics{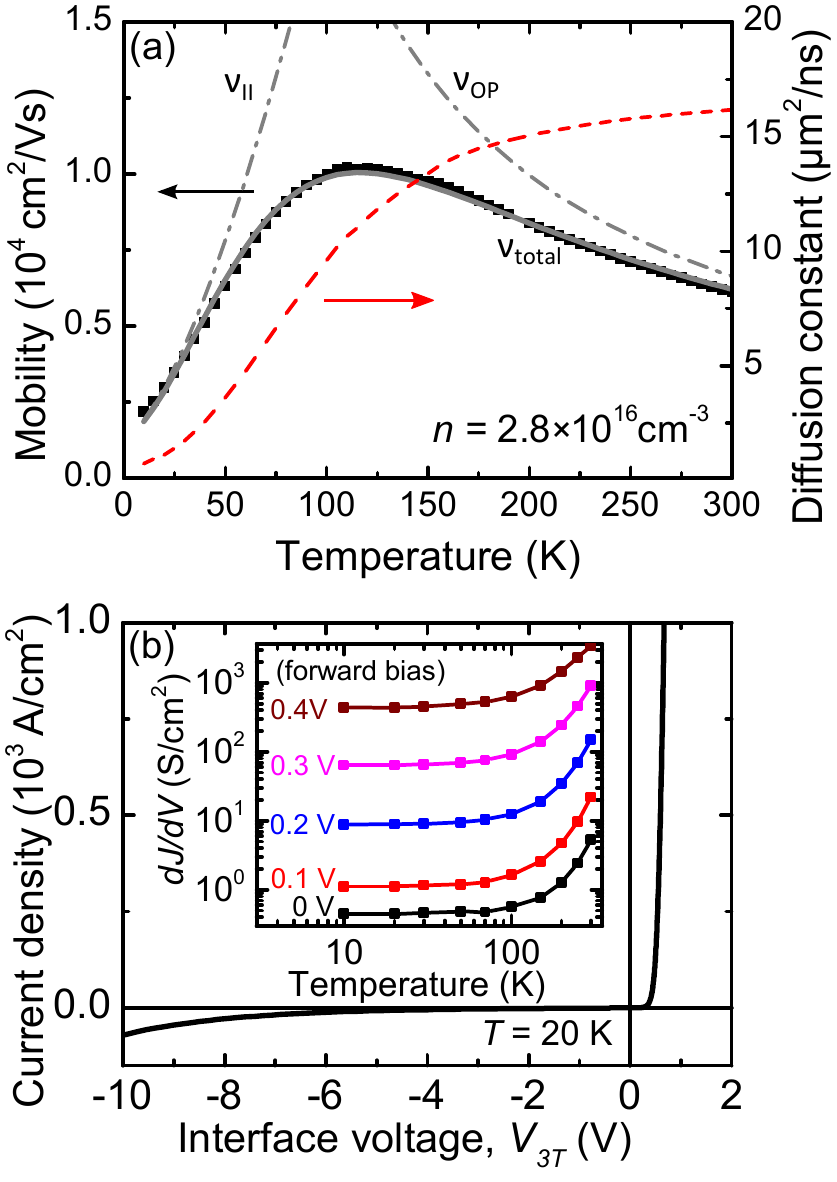}
\caption{(Color online) (a) The \textit{n}-GaAs mobility extracted
from Hall measurements (left ordinate) as a function of
temperature on the Co$_2$FeSi heterostructure. The gray solid line
is a fit to the model for the mobility given by
Eq.~\ref{mobility}, with the ionized-impurity (II) and
optical-phonon (OP) scattering contributions to the mobility
indicated with the dash-dot gray lines. In the fit shown,
$A=1.3\times 10^3$~cm$^2$V$^{-1}$s$^{-1}$,
$B=18$~cm$^2$V$^{-1}$s$^{-1}$K$^{-3/2}$ and $C=2.0\times
10^6$~cm$^2$V$^{-1}$s$^{-1}$K$^{-1}$. The red dashed line (right
ordinate) is the channel diffusion constant calculated with
Eq.~\ref{Einstein}. (b) Typical Co$_2$FeSi contact 3-terminal
$J-V$ characteristic at 20 K. The inset in (b) is the differential
conductance as a function of temperature at different interface
forward bias voltages. The solid curves connect data points.}
\label{fig:chargetransport}
\end{figure}

Standard multiprobe dc transport measurements were performed as a
function of temperature to characterize both the \textit{n}-GaAs
channel and the Co$_2$FeSi/\textit{n}-GaAs interface. A companion
Hall bar was fabricated from the same heterostructure used to
fabricate the SV devices, and transport measurements were
performed from 10-350 K to extract the carrier concentration and
mobility of the \textit{n}-GaAs. The Hall carrier concentration
was measured to be $2.8\times 10^{16}$ cm$^{-3}$ for the
Co$_2$FeSi heterostructure and $3.5\times 10^{16}$ cm$^{-3}$ for
the Co$_2$MnSi heterostructure. Fig.~\ref{fig:chargetransport}(a)
shows the channel electron mobility and diffusion constant as a
function of temperature for the Co$_2$FeSi heterostructure. The
Hall factor \cite{Yu1996}, which causes deviation of the Hall
mobility from the electron mobility in \textit{n}-GaAs, is
accounted for by assuming the Hall factor is unity at 300 K
\cite{Mazuruk1986,Lewis1955} and that the carrier concentration is
temperature-independent.

A typical SV device Co$_2$FeSi/\textit{n}-GaAs contact
three-terminal (3T) interface current-voltage ($J-V$)
characteristic is shown in Fig.~\ref{fig:chargetransport}(b). The
inset of Fig.~\ref{fig:chargetransport}(b) shows the differential
conductance per unit area ($dJ/dV$) as a function of temperature.
Tunneling-dominated transport (field emission) is known to be
necessary for spin injection in FM/GaAs Schottky contacts
\cite{Hanbicki2003}. The existence of tunneling-dominated
transport under forward bias at all temperatures is supported by
two observations. First, $dJ/dV$ increases exponentially with
forward bias voltage at all temperatures, at a rate that is
independent of temperature. Because of the triangular Schottky
barrier \cite{Cowley1965}, the forward bias voltage across a
Schottky interface changes the thickness of the effective
potential barrier through which tunneling occurs
\cite{Burstein_Lundqvist, Brinkman1970}. Although thermionic
emission and thermionic field emission also lead to an exponential
increase of $dJ/dV$ with interface forward bias voltage, the rate
for those processes is strongly temperature-dependent, ruling out
those mechanisms. Second, at temperatures below the Fermi
temperature of the \textit{n}-GaAs ($\sim$~60 K for these samples)
the forward bias differential conductance decreases weakly with
decreasing temperature. Although $dJ/dV$ at forward bias is
temperature-dependent above the Fermi temperature, this does not
imply thermionic emission but rather an increase in the tunneling
attempt rate due to the nondegeneracy of the \textit{n}-GaAs
\cite{Burstein_Lundqvist}.

\subsection{Spin transport}
A schematic diagram of the SV measurement is shown in
Fig.~\ref{fig:diagram}(a). A dc bias current $J_i$ flows through
the injector contact and a second bias current $J_d$ flows through
the detector contact. The injector and detector current sources
share a common remote reference contact. In this article positive
currents and interface voltages refer to electron extraction from
the channel, \textit{i.e.,} forward bias of the
metal/semiconductor Schottky contact. The bias current applied to
the detector contact results in a voltage drop $V_{d}$ over the
tunnel barrier, which is the 3T interface voltage of the detector
contact. In these devices, a forward bias applied at the detector
contact enhances the nonlocal SV signal size compared to an
unbiased detector (zero detector bias is the traditional nonlocal
SV configuration pioneered by \citet*{Johnson1985}). We will
henceforth refer to the case of a bias current applied through the
detector contact as the biased-detector spin valve (BDSV)
measurement. The enhancement in the SV signal size with a bias
applied to the detector contact has been observed in prior
\textit{n}-GaAs lateral SV literature on similar heterostructures
\cite{Crooker2009, Bruski2013}, and the possible origins will be
discussed in detail later in this article.

An applied magnetic field is swept along the FM easy axis to
switch the magnetizations of the injector and detector contacts
from the parallel to antiparallel configuration, which allows for
a definitive measurement of the nonlocal voltage due to spin
accumulation. The difference in the nonlocal detector voltage
$V_{nl}$ between the parallel and antiparallel contact
magnetization states is due to spin accumulation in the
semiconductor \cite{Johnson1985} and is given by
\begin{equation}\label{Vnl}
\Delta V_{nl} = V_{NL,\uparrow \uparrow}-V_{NL,\uparrow
\downarrow} = \eta(V_d) \frac{n_{\uparrow}-n_{\downarrow}}{e}
\frac{\partial \mu}{\partial n},
\end{equation}
where $n_{\uparrow (\downarrow)}$ is the majority (minority)
spin-resolved carrier density in the GaAs channel, $e$ is the
electron charge, and $\partial \mu/\partial n$ is the inverse of
the thermodynamic compressibility of the semiconductor. We will
refer to $n_{\uparrow}-n_{\downarrow}$ as the spin accumulation
and $(n_{\uparrow}-n_{\downarrow})/n$ as the dimensionless spin
polarization throughout this article. The dimensionless detection
efficiency parameter $\eta(V_d)$ characterizes the spin
sensitivity of the detection contact \cite{Song2010} and is a
function of the bias voltage. Because of the bias current applied through the detector contact,
$V_{nl}$ is not an open circuit nonlocal voltage (or
``electromotive force"). The voltage drop over the detector
Schottky tunnel barrier contributes an offset $V_{d}$, so that
\begin{equation}\label{VBD}
V_{nl} = V_{d} + \frac{\Delta V_{nl}}{2} \hat{\mathbf{m}}_i \cdot
\hat{\mathbf{m}}_d
\end{equation}
where $\hat{\mathbf{m}}_{i(d)}$ is the unit vector specifying the
magnetization of the injector (detector) contact.

Example BDSV field sweeps are shown in Figs.~\ref{fig:diagram}(b)
and (c) on SV devices with an injector-detector edge-to-edge
separation of 250 nm at an injector bias current of $J_i = 10^3$
A/cm$^2$. The BDSV measurement on the device with Co$_2$FeSi
contacts is shown in Fig. \ref{fig:diagram}(b) at $J_d = 40$
A/cm$^2$, and for the device with Co$_2$MnSi contacts in Fig.
\ref{fig:diagram}(c) at $J_d = 400$ A/cm$^2$. The
Co$_2$MnSi/\textit{n}-GaAs contacts exhibited large voltage noise
in the nonlocal SV measurements, and the signal-to-noise ratio
(SNR) was not adequate for measurements at high temperatures. For
this reason, the analysis presented in this article is carried out
for measurements on Co$_2$FeSi/\textit{n}-GaAs devices. At low
temperatures, at which the SNR in Co$_2$MnSi/\textit{n}-GaAs
devices was adequate, the SV measurements were quantitatively
similar to those on Co$_2$FeSi/\textit{n}-GaAs devices. A linear
background in $V_{nl}$ can result from the Hall effect due to
slight misalignment. The slope, which is a weak function of
temperature, is subtracted from the data before extracting $\Delta
V_{nl}$.

Nonlocal Hanle (NLH) measurements \cite{Johnson1985, Jedema2002}
were also performed in the biased-detector configuration. In the
NLH measurement a magnetic field applied perpendicular to the
sample plane is used to apply a precessional torque, which, in
combination with diffusion, dephases the spin accumulation. In all
of the NLH measurements, the applied field was small enough so
that the out-of-plane rotation of the contact magnetization
decreased the in-plane component of the magnetization by less than
1.5\%, which was considered negligible. The NLH measurement could
be executed with the injector and detector contacts in either the
parallel or antiparallel configuration. In the fitting of the NLH
lineshape discussed in Section \ref{Hanle}, the difference of the
parallel and antiparallel field sweeps is used.

At cryogenic temperatures, the NLH measurement in \textit{n}-GaAs
is complicated by the strong hyperfine fields due to dynamic
nuclear polarization (DNP) \cite{Chan2009, Salis2009,
MeierZak1984}. Steady-state conditions are difficult to achieve
due to long ($\sim$ seconds) nuclear depolarization timescales,
and small misalignments between the applied field and the contact
magnetization result in oblique Overhauser fields, which distort
the NLH lineshape \cite{Chan2009, Salis2009}. To mitigate the
influence of DNP effective fields on the NLH lineshape, a low duty
cycle ($< 1 \%$) pulsed current measurement was used for the NLH
sweeps at temperatures below 100 K. The current was turned off for
1000 milliseconds, then pulsed on for 5 milliseconds after which
the voltage was recorded and the pulse-train repeated. The current
rise and fall times were much shorter than the few-millisecond
current pulse duration. The pulsed measurement minimizes the
nuclear polarization buildup because the current is on for a time
much less than the nuclear polarization time \cite{MeierZak1984}.
Example NLH data obtained for the 250 nm separation Co$_2$FeSi
device at 60 K are shown in Fig. \ref{fig:diagram}(b).

\section{Results}

\subsection{Effect of detector bias}
We now discuss the effect of detector bias on our SV measurements.
First, we note that \citet{Crooker2009} and \citet{Bruski2013}
observed similar enhancement of the spin valve signal in the
presence of a detector bias current or voltage. Although several
mechanisms have been proposed to explain the enhancement in the
nonlocal SV signal with detector bias, the enhancement remains
poorly understood. At the end of this section, we will return to
discuss possible explanations in light of our measurements.

\begin{figure}
\includegraphics{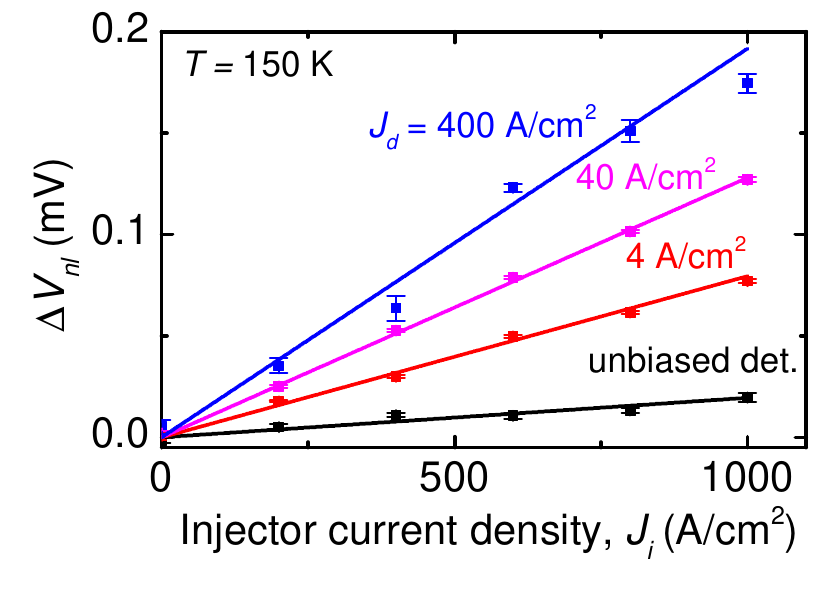}
\caption{(Color online) Injector bias current dependence of
$\Delta V_{nl}$, for varying detector forward bias currents, on
the 250 nm separation device at 150 K. The lines shown are linear
fits. }\label{fig:ibias}
\end{figure}

We find that a sufficiently large forward bias current applied
through the detector contact increases the SV signal $\Delta
V_{nl}$ at all temperatures. Fig. \ref{fig:ibias} shows $\Delta
V_{nl}$ vs. $J_i$ for the 250 nm separation at 150 K. $\Delta
V_{nl}$ increases linearly with $J_i$ at all detector bias
currents, but the slope of $\Delta V_{nl}$ vs. $J_i$ is enhanced
with increasing detector forward bias current. This enhancement is
particularly advantageous for measurements at high temperatures
near 300 K, at which the spin valve signal becomes small in
\textit{n}-GaAs \cite{Lou2006a, Saito2013}.
This effect was observed in devices with both Co$_2$FeSi and
Co$_2$MnSi contacts and was observed previously for devices with
Fe contacts \cite{Crooker2009}.

\begin{figure}
\includegraphics{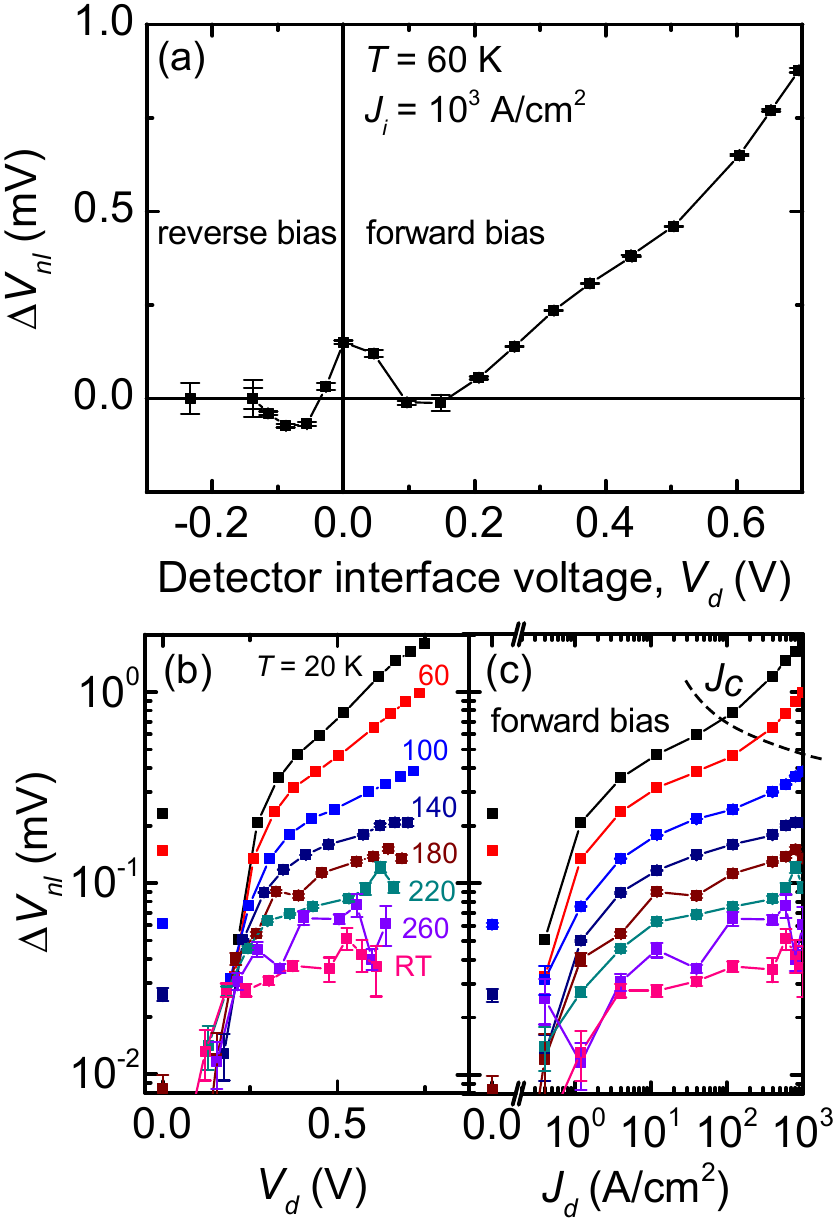}
\caption{(Color online) (a) $\Delta V_{nl}$ as a function of
detector interface voltage $V_d$ for fixed injector bias current.
(b,c) The detector forward bias \textit{voltage} (b) and
\textit{current} (c) dependence of $\Delta V_{nl}$ from 20 K to
room temperature (RT). Only the zero detector bias and forward
bias points are shown in (b) and (c) to illustrate the enhancement
of $\Delta V_{nl}$ at forward detector bias. The dashed line in
(c) indicates $J_c$, above which spin drift in the channel caused
by the detector bias current enhances the spin accumulation at the
detector. All data shown in this figure were taken with the 250 nm
injector-detector separation device, and $J_i=10^3$~A/cm$^2$.}
\label{fig:BDSV}
\end{figure}

For the case of no bias current passing through the detector ({\it
i.e.} the conventional nonlocal SV measurement), $\Delta V_{nl}$
could be measured in the 250 nm separation device for temperatures
less than approximately 200 K (see data points in Fig.
\ref{fig:BDSV}(b-c) at $V_{d} = J_d = 0$). For a fixed injector
current, the SV measurement was then performed at different
detector bias currents. The corresponding interface voltage drop
$V_d$ was measured at each bias current, and so the data may be
presented as a function of either bias voltage $V_d$ or current
$J_d$. The results of this measurement at 60 K on the 250 nm
separation are shown in Fig. \ref{fig:BDSV}(a) and are summarized
for all temperatures in Figs. \ref{fig:BDSV}(b) and (c). At
forward detector bias above interface voltages of $V_d \sim$0.2 V,
we observe significant enhancement of $\Delta V_{nl}$. As shown in
Fig. \ref{fig:BDSV}(a), the dependence of $\Delta V_{nl}$ on the
detector bias is non-monotonic below $\sim$200~K, and it is
suppressed at small detector voltages (of either sign) and even
changes sign for a narrow window of reverse bias. Although
$V_{nl}$ is sensitive to 3T signals \cite{Lou2006b} produced by
\textit{local} spin injection at the detector contact, only
\textit{nonlocally}-injected spin accumulation contributes to
$\Delta V_{nl}$ in a spin valve measurement, because $\Delta
V_{nl}$ is the difference in nonlocal voltage between parallel and
antiparallel magnetization states. Furthermore, as shown in Fig.
\ref{fig:diagram}(b), the NLH measurement can also be performed
with the parallel-antiparallel difference at zero field matching
the BDSV magnitude. The existence of the NLH effect at low
temperatures demonstrates conclusively that the biased-detector
measurement in these devices is a probe of the nonlocally injected
spin accumulation.

The enhancement in $\Delta V_{nl}$ under forward detector bias
occurs at all temperatures measured, from 20 K to room
temperature. Using the BDSV measurement a clear SV signal could be
measured on the separations below 1 $\mu$m up to and above room
temperature on the Co$_2$FeSi devices. To our knowledge, the spin
signal we measure on the 250 nm separation device of $\sim$40
$\mu$V at room temperature is over an order-of-magnitude larger
than that which has been achieved in FM/\textit{n}-GaAs SVs, to
date \cite{Saito2013}. We now discuss the possible origins of the
forward bias enhancement of the SV signal.

We consider first the influence of drift due to electric fields in
the channel between the injector and detector contacts. Due to the
relatively low carrier density in these samples, the spin drift
length $l = \tau_s J/ne$ can be comparable to or larger than the
spin diffusion length $\lambda = \sqrt{D \tau_s}$
\cite{Yu2002,Tahara2016}. In the case of a forward bias current
applied through the detector contact (electron extraction from the
channel), the electric field in the channel causes drift of
electrons from the injector towards the detector contact,
enhancing the nonlocal spin accumulation when compared to spin
diffusion alone. To determine if the detector bias current leads
to significant drift enhancement of $\Delta V_{nl}$, the current
density in the channel between injector and detector contacts at
which the spin drift length was equal to the spin diffusion length
was evaluated at each temperature. Above a critical current
density $J_{c} = n e \sqrt{D/\tau_s}$, which is the current
density at which $l=\lambda$, drift enhancement of the nonlocal
spin accumulation below the detector contact becomes significant.
The region where this occurs is illustrated in Fig.
\ref{fig:BDSV}(c), in which the dashed curve shows $J_{c}$. The
drift enhancement is significant only at low temperatures and the
highest detector bias currents. This is in contrast to the case of
Si described in Ref. \cite{Tahara2016} in which the long spin
lifetime at room temperature, combined with higher current
densities than we apply, leads a spin drift length which can be
much longer than the spin diffusion length. Because the
enhancement in $\Delta V_{nl}$ occurs at all temperatures and for
current densities far below $J_c$, it cannot be attributed solely
to spin drift effects in the channel. Although variations on
simple drift models have been proposed \cite{Chantis2008}, it is
unlikely that drift alone can play a significant role given that
the enhancement is observed up to room temperature.  For the
purposes of discussion, we attribute the enhancement in $\Delta
V_{nl}$ with detector forward bias primarily to enhancement of
$\eta$, the detection efficiency, which we treat as a purely
interfacial property. The detection efficiency is a function of
detector bias, \textit{i.e.} $\eta \rightarrow \eta(V_d)$.

\citet{Hu2011} and \citet{Salis2010} observed a highly
non-monotonic behavior of the sign of the injected spin
polarization in similar heterostructures with Fe contacts. The
sign and magnitude depended strongly on the details of the
\textit{n}-GaAs band structure in the region of $n^+$ doping near
the interface. It is possible that the enhancement of $\eta$ under
forward bias is due to the enhanced participation of additional
quantum well states that form on the SC side of the tunnel barrier
due to the $n^+$ doping layer. It has been proposed that these
states play a critical role in both charge and spin current in
tunnel contacts using Schottky barriers through FM/SC
wavevector-matching arguments which depend on the degree of
quantum confinement of the SC states \cite{Dery2007}.

Another point of view focuses on the nonlinear current-voltage
characteristic of the tunnel barrier itself
\cite{Pu2013,Shiogai2014}. A simple analysis suggests that the
ratio of the detected voltage to the spin accumulation should be
modified by the ratio $(J/V)/(dJ/dV)$ of the absolute to
differential conductance, although \citet{Jansen2015} have noted
that this correction factor is in fact an upper bound.  In our
case, however, we observe an effect that is opposite to that
suggested by this argument. $(J/V)/(dJ/dV)$ is smaller at forward
bias voltage than at zero bias, because $J$ increases
exponentially with $V$.

Because the bias current applied to the detector introduces a 3T
offset $V_d$ to $V_{nl}$, care must be taken to separate signals
due to nonlocal spin accumulation from signals of local origin.
Surface localized states in tunnel barriers have been at the
center of a controversy in the semiconductor spin injection
literature because of the influence these states can have on both
the magnitude and lineshape of the 3T Hanle measurement
\cite{Tran2009}. For example, \citet{Txoperena2014} determined
that impurity-assisted tunnelling processes can lead to
Lorentzian-shaped magnetoresistance effects that mimic the Hanle
effect. Also, \citet{Jansen2012} note that in the 3T geometry the
change in 3T voltage due to spin accumulation can originate from
spin accumulation in interface localized states as well as bulk
channel spin accumulation. Our measurement, however, probes the
parallel-antiparallel difference in the nonlocal voltage,
notwithstanding the bias applied to the detector contact. Although
localized states may play an important role in the spin-polarized
transport at our interfaces, the mechanisms discussed by
\citet{Txoperena2014, Jansen2012} are only relevant for 3T local
spin detection where the ferromagnetic contact simultaneously
serves as the injector and detector.

Another possible physical explanation for the detector bias
dependence of $\Delta V_{nl}$ is that significant features exist
in the spin-resolved density-of-states (DOS) of the
Co$_2$FeSi/GaAs interface near the Fermi level. These features
could lead to spin injection and detection efficiencies that vary
with forward bias voltage, as states above the Fermi level in the
FM become available for elastic tunnelling from the SC. Density
functional theory (DFT) calculations done for Co$_2$FeSi in the
L2$_1$ phase\cite{Wurmehl2005, Balke2006a} suggest strong
variations in the bulk minority DOS near the Fermi level over
energy ranges of $\sim$hundreds of meV, which are comparable to
the scale of the interface voltages at the detector in our
measurement. Strong bulk minority DOS variations near the Fermi
level have also been predicted for Co$_2$MnSi which are largely
insensitive to the phase (L2$_1$ vs. B2)\cite{Picozzi2007}.
However, the bias dependence of spin detection shown in Fig.
\ref{fig:BDSV}(a) cannot be clearly correlated with the features
in the spin-resolved DOS reported by DFT calculations.
Additionally, interface states, such as those which have been
proposed for the Fe/GaAs(001) interface, will contribute to the
tunneling current\cite{Chantis2007}. Although it is likely that
the low-voltage features in $\Delta V_{nl}(V_d)$ are associated
with electronic structure of the interface, we have no
quantitative description of the bias-dependence of the nonlocal
voltage.

We now comment briefly on the sign of the spin valve signals we
observe. In this article, a decrease in $V_{nl}$ in the
antiparallel magnetization state is defined as a positive $\Delta
V_{nl}$. The BDSV sweeps shown in Figs. \ref{fig:diagram}(b) and
(c) are examples of positive $\Delta V_{nl}$ values. The sign of
$\Delta V_{nl}$ is determined by the relative signs of the
injection and detection efficiencies. That is, same sign (opposite
sign) injection and detection efficiencies correspond to a
positive (negative) $\Delta V_{nl}$. Microscopically, the
individual signs of these efficiencies are determined by the
difference in the spin-resolved interface conductances
$g_{\uparrow}-g_{\downarrow}$, where the ``up" direction is
defined by the energy-integrated majority spin direction
(\textit{i.e.}, magnetization) of the ferromagnet.  Because the
nonlocal voltage depends on the product of the two efficiencies,
it is not possible to correlate its sign directly with the sign of
the spin accumulation.  At low temperatures, the influence of the
electronic Knight field on the nuclear polarization in oblique
Hanle geometries \cite{MeierZak1984, Chan2009} can be used to
determine the sign of the spin accumulation with respect to the
magnetization orientation. We have determined that at high forward
bias (spin extraction) the sign of the spin accumulation is
minority in Co$_2$FeSi and majority in Co$_2$MnSi with respect to
the magnetization of the injector contact \cite{Christie2014}.
\subsection{Injector-detector separation dependence}
\begin{figure}
\includegraphics{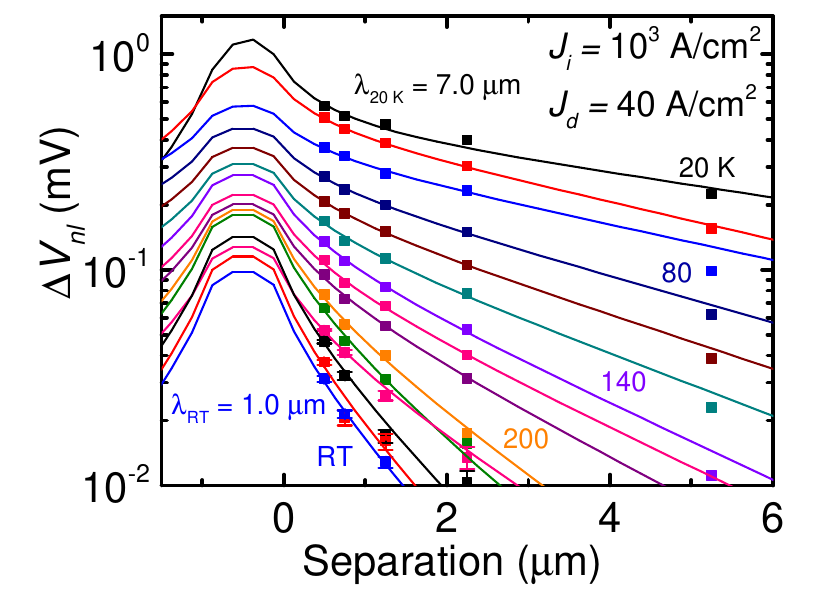}
\caption{(Color online) The injector-detector separation
dependence of $\Delta V_{nl}$ for the devices with Co$_2$FeSi
contacts at temperatures from 20 K to 300 K, in increments of 20
K. The horizontal axis of the plot is the injector edge to
detector center separation, \textit{i.e.} the 1 $\mu$m-wide
injector extends from -1 to 0 $\mu$m on the horizontal axis.
Superimposed as solid lines are the fits of a 2D numerical
solution of Eq.~\ref{DD} with $\tau_s$ and $\eta\alpha$ as the
fitting parameters. The bias conditions are indicated on the
figure as well as the spin diffusion lengths at 20 K and room
temperature (RT). At low temperature, the IDSD measurement on the
Co$_2$MnSi devices yielded comparable SV signal sizes and
\textit{n}-GaAs spin diffusion length. A complete
temperature-dependence measurement, however, was not performed.}
\label{fig:sepdep}
\end{figure}

We quantify device parameters at different temperatures using the
injector-detector separation dependence (IDSD) of the spin valve
signal size, rather than relying on NLH measurements. The NLH
measurement in \textit{n}-GaAs becomes challenging at high
temperatures because of the magnetoresistance backgrounds present
over the much larger magnetic field range required when the spin
lifetime is small. The injector-detector separation was varied in
order to extract the spatial dependence of the spin accumulation
in the channel. By utilizing the enhanced signal in the BDSV
configuration, clear SV signals could be measured at the smallest
separations up to room temperature. For the IDSD measurement, the
detector contact forward bias was fixed at a current density of 40
A/cm$^2$. This bias current was well into the enhancement regime
shown in Fig. \ref{fig:BDSV}(c), but below the regime where spin
drift enhancements were significant at low temperatures. $\Delta
V_{nl}$ was recorded at bias conditions $J_i= 1000$ A/cm$^2$, $J_d
= 40$ A/cm$^2$ for each temperature and injector-detector
separation. The results of the IDSD measurement are summarized in
Fig. \ref{fig:sepdep}. The solid lines in Fig. \ref{fig:sepdep}
are fits to a numerical model of the spin accumulation in the
channel, which will be explained in detail later in this article.

We note that in Eq.~\ref{Vnl}, $\Delta V_{nl}$ is proportional to
the spin accumulation $n_{\uparrow}-n_{\downarrow}$ and the
inverse compressibility of the channel $\partial \mu/\partial n$.
At temperatures above the Fermi temperature (in our samples $T_F
\simeq 60$~K) at which the \textit{n}-GaAs is no longer
degenerate, $\partial \mu/\partial n$ is a function of
temperature. In the nondegenerate regime ($T \gg T_F$ ), $\partial
\mu/\partial n \propto T$. This relationship implies that as the
temperature increases in the nondegenerate regime, a larger
$\Delta V_{nl}$ is measured \textit{for a given spin
accumulation}. For these samples,
\begin{equation}
\frac{\partial \mu}{\partial n}\bigg|_\text{300 K} \simeq
7\frac{\partial \mu}{\partial n}\bigg|_\text{20 K}.
\end{equation}
Because of this enhancement factor, while the spin accumulation
falls by two orders of magnitude from 20~K to 300~K, $\Delta
V_{nl}$ at separations much smaller than a diffusion length only decreases by roughly one order of magnitude over
the same temperature range.

\subsection{Modeling of the spatial decay of spin accumulation}
Here we discuss the model used to describe the spin accumulation
in the channel and which is used to fit the IDSD measurement results.
Typically, in systems where spin diffusion is one-dimensional, the
SV signal size is interpreted with the expression
\cite{Johnson1985}
\begin{equation}\label{Rnl}
\Delta R_{nl} = \Delta V_{nl}/I = \frac{\eta^2 \rho \lambda
e^{-y/\lambda}}{A},
\end{equation}
where $\rho$ is the channel resistivity, $A$ is the channel
cross-sectional area, and $y$ is injector-detector separation.
Eq.~\ref{Rnl} has been used to model the SV signal size in a
variety of material systems \cite{Jedema2002,
Lou2006a,Tombros2007} in which the FM/NM barrier resistance is
much larger than the channel spin resistance, so that the
conductivity mismatch problem \cite{Rashba2000} may be ignored. We
choose to use a more general numerical model of the spin
accumulation in the channel to fit to the IDSD measurement because
of several considerations. First, as discussed earlier, drift due
to the bias current influences the spatial spin accumulation
profile in \textit{n}-GaAs at low temperatures, and the exact
drift field is best captured by a numerical model. Second, at
measurement temperatures near room temperature the spin diffusion
length in \textit{n}-GaAs is less than the channel thickness of
2.5 $\mu$m. In this regime a more general solution of the spin
drift-diffusion equation is needed, because Eq.~\ref{Rnl} is only
appropriate for devices where the spin drift and diffusion are
effectively one dimensional. In two or three dimensions, the spin
accumulation decays faster than $e^{-y/\lambda}$ for $y <
\lambda$, in exact analogy to the two and three dimensional
solutions of the screened Poisson equation.

The spatial profile of spin accumulation in the channel is
modeled by solving the spin drift-diffusion equation
\cite{Yu2002} in steady state,
\begin{equation}\label{DD}
\frac{\partial \mathbf{P}}{\partial t} = 0 =
-\frac{\mathbf{P}}{\tau_s}+D \nabla^2 \mathbf{P} +
\frac{\mathbf{J}}{n e}\cdot\nabla \mathbf{P} + \frac{\alpha
\hat{\mathbf{m}}_{i} |\mathbf{J}_i|}{n e \Delta z},
\end{equation}
where $|\mathbf{P}| \equiv (n_{\uparrow}- n_{\downarrow})/n$ is
the dimensionless spin polarization of the channel, $D$ is the
spin diffusion constant (equal to the charge diffusion constant
\cite{Yu2002}), $\hat{\mathbf{m}}_{i}$ specifies the injector
contact magnetization direction, and the last term specifies the
source term, which is only nonzero at the cells of the finite
element model where spin injection occurs. In the source term, the
$\Delta z$ factor in the denominator is the size of the injection
cell in the $z$-direction, which normalizes the injection rate in
the finite-element grid properly. $\mathbf{J}$ is the current
density in the channel, and the parameter $\alpha$ is the spin
injection efficiency at the FM/SC interface (\textit{i.e.} for
$\alpha = 1$ the spin current at the FM/SC interface is equal to
the charge current). $\alpha$ encompasses both the bulk
polarization of the current in the FM, as well as interface
effects determining the polarization of the charge current.
\begin{figure}
\includegraphics{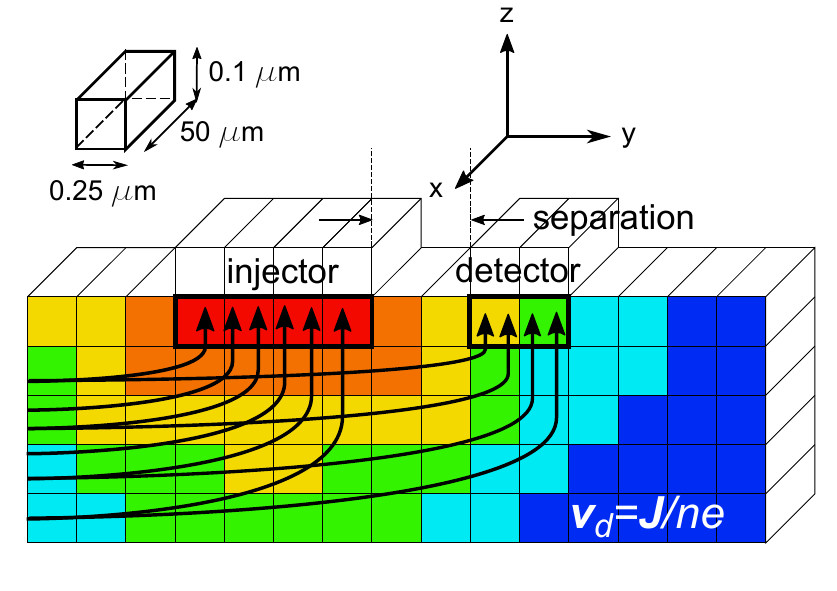}
\caption{(Color online) Schematic illustrating the 2D
finite-element model used to solve Eq.~\ref{DD} numerically. The
spin accumulation, which drifts and diffuses from the injector
contact, is indicated for illustrative purposes in false color
(red high, blue low). The channel drift velocity $\mathbf{v}_d =
\mathbf{J}/ne$ is schematically shown by the field lines. The
bolded black outlines the cells in which injection and detection
occurs. The cell dimensions ${\Delta x, \Delta y, \Delta z}$ used
in the simulation are shown in the upper left. The number of cells
drawn is not the actual number of cells used, nor is the model
drawn to scale.} \label{fig:model}
\end{figure}
The spin valve device geometry is cast into a finite-element grid,
and Eq.~\ref{DD} is solved numerically by forward iteration until
steady state is reached. See Fig. \ref{fig:model} for a schematic
diagram illustrating the model geometry. The contact length in the
$x$-direction (50 $\mu$m) is much longer than the spin diffusion
length at all temperatures. The model is therefore confined to the
$yz$-plane and the spin accumulation is assumed to be uniform in
the $x$-direction. Neumann boundary conditions are enforced at the
free boundary cells, \textit{i.e.} the diffusive spin current
$\propto \nabla \mathbf{P} = 0$ at the boundaries.

The current density $\mathbf{J}$ in the channel was solved for
prior to solving Eq.~\ref{DD} by assuming charge neutrality
throughout the channel, so that $\nabla \cdot \mathbf{E} = \nabla
\cdot \mathbf{J} = 0$. Because $\nabla \cdot \mathbf{J} = 0$,
there exists a scalar potential $\phi_J$ that satisfies $\nabla^2
\phi_J = 0$. $\phi_J$ is solved for with a Laplace relaxation
method, and finally the current density vector field is solved for
by evaluating $\nabla \cdot \phi_J = \mathbf{J}$.

The diffusion constant $D$ is calculated from the Einstein relation
\begin{equation}\label{Einstein}
D = \frac{n \nu}{e}\left(\frac{\partial \mu}{\partial n}\right),
\end{equation}
where $\nu$ is the mobility. For $n = 2.8\times 10^{16}$ GaAs, the
Fermi temperature $T_F\simeq $60 K, so in order to capture the
transition from degenerate to nondegenerate behavior, the inverse
compressibility $\partial \mu /\partial n$ is calculated using
full Fermi-Dirac statistics. A parabolic conduction band density
of states with GaAs effective mass $m^*=0.067m_0$ \cite{Yu1996} is
used, and the inverse compressibility is evaluated via the
expression
\begin{equation}\label{compress}
\frac{\partial \mu}{\partial n} = \frac{k_b
T}{n}\frac{F_{1/2}(\zeta)}{F_{-1/2}(\zeta)},
\end{equation}
where $\zeta \equiv \mu / k_b T$ is the reduced chemical potential
and $F_\alpha(\zeta)$ is the complete Fermi-Dirac integral. In the
limits $T \ll T_F$ and $T \gg T_F$ Eq.~\ref{compress} reduces to
$\partial \mu/\partial n = 2 E_{\text{F}} / 3 n$ and $\partial
\mu/\partial n = k_b T / n$, respectively.

To compare the solution of Eq.~\ref{DD} directly with the measured
$\Delta V_{nl}$, the calculated nonlocal spin accumulation at the
detector is input to Eq.~\ref{Vnl}. The overall scale of $\eta$,
the detection efficiency, cannot be determined in this
measurement. However, because the known injector current density
constrains the spin injection rate, the product of the injection
and detection efficiencies $\eta \alpha$ can be determined. We
will discuss the constraints on $\eta$ in more detail below.

\begin{figure}
\includegraphics{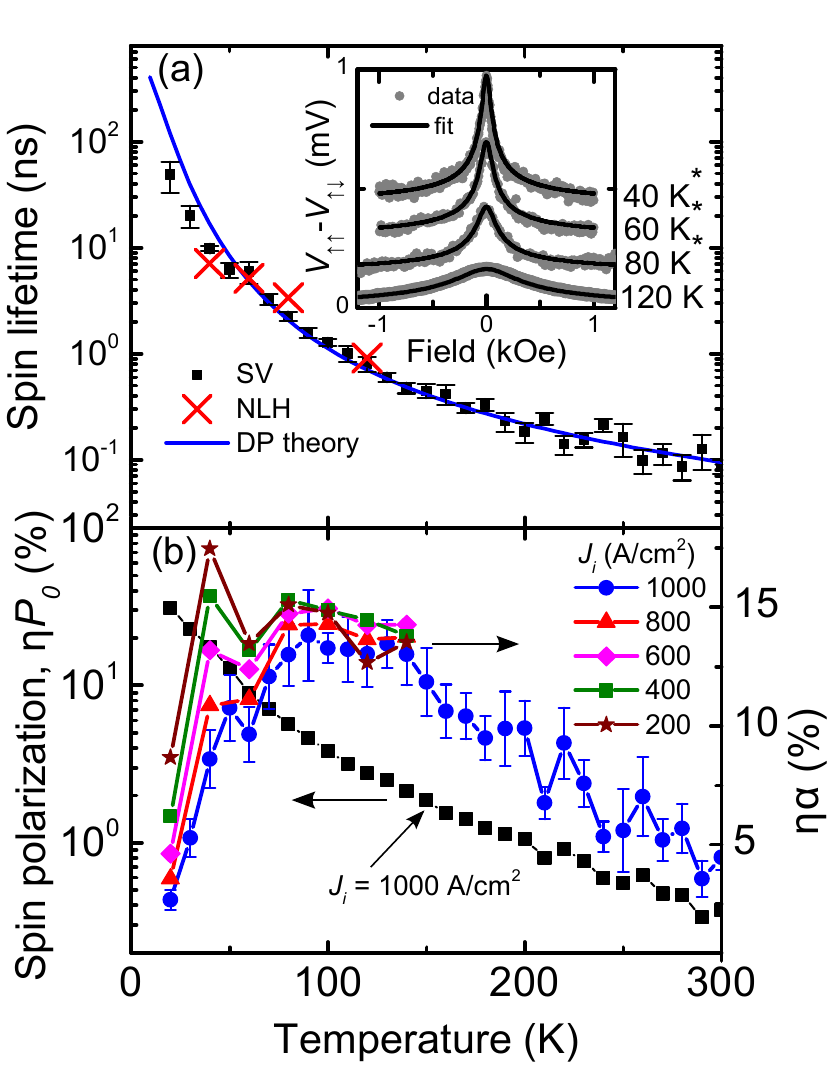}
\caption{(Color online) (a) The temperature dependence of $\tau_s$
extracted from the fits in Fig. \ref{fig:sepdep} along with the
theoretical prediction based on Eq.~\ref{taus}, which is shown as
the blue solid line. Spin lifetimes extracted from NLH
measurements are shown as red crosses, with the corresponding NLH
data $V_{\uparrow \uparrow}-V_{\uparrow \downarrow}$ and fits to
Eq.~\ref{NLH} shown in the inset (artificially offset). The
asterisks on the temperature labels in the inset indicate that the
NLH sweeps were taken with the pulsed current measurement to
mitigate DNP effects. The NLH data shown are taken at the same
bias currents as used for the data of Fig. \ref{fig:sepdep} on the
250 nm separation device. (b) The temperature dependence of $\eta
P_0$ (left ordinate) and $\eta\alpha$ (right ordinate). $P_0$ is
the spin polarization directly beneath the injector from the model
fits shown in Fig. \ref{fig:sepdep}. At temperatures below 140 K,
$\eta\alpha$ is shown for different injector current densities
using the symbols indicated in the legend. In (b) representative
error bars are shown for the $J_i = 10^3$~A/cm$^2$ data only. All
data in (b) were taken with $J_d = 40$~A/cm$^2$. }
\label{tauetaalpha}
\end{figure}

The IDSD measurement results are fit to the numerical solution of
Eq.~\ref{DD}, with the spin lifetime $\tau_s$ and the
dimensionless spin injection efficiency $\alpha$ as fitting
parameters. The fits to the IDSD results are shown as solid lines
in Fig. \ref{fig:sepdep}, and the temperature dependence of the
fitting parameters $\tau_s$ and $\eta \alpha$ are shown in Figs.~
\ref{tauetaalpha}(a) and (b). The product $\eta P_0$ of the
detection efficiency and the spin polarization $P_0$ below the
injector is also shown in Fig.~\ref{tauetaalpha}(b).

\subsection{Hanle fitting} \label{Hanle}

At low temperatures, at which the NLH measurement could be
performed, the spin lifetime obtained from fits of the IDSD
measurement could be compared to the spin lifetime measured by
Hanle precession experiments. To fit NLH field sweeps the data
were fit to the Green's function solution of Eq.~\ref{DD} in one
dimension, which gives
\begin{equation}\label{NLH}
V_{nl}(H) \propto \mathbf{P}(y)\cdot \hat{\mathbf{m}}_{d} \propto
\int_{-\infty}^t \frac{\text{exp}[-(\frac{y^2}{4 D t}
+\frac{t}{\tau_s})]}{\sqrt{4 \pi D t}} \cos(\gamma_e H t) dt,
\end{equation}
where $|\gamma_e|/2\pi=$ 0.62~MHz/Oe is the gyromagnetic ratio in
GaAs. Eq.~\ref{NLH} is identical to solving Eq.~\ref{DD} in one
dimension with an added precession term from an external
transverse magnetic field $H$, and $\mathbf{J} = 0$. The
simplification to one dimension is appropriate at low
temperatures, because the spin diffusion length $\sqrt{D \tau_s}$
is larger than the channel depth of 2.5~$\mu$m.

\subsection{Spin lifetime calculation}
In order to compare the measured temperature dependence of the spin
lifetime with DP theory, we used the method of \citet*{Lau2001,
Lau2004} to calculate the spin relaxation rate for the doping
concentration \textit{n}$ = 2.8 \times 10^{16}$ cm$^{-3}$. The
spin relaxation rate, $\tau_s^{-1}$, can be expressed as
\begin{equation}\label{taus}
\tau_s^{-1} = \frac{1}{\tilde{n}}\int
D(\epsilon)f(\epsilon)[1-f(\epsilon)]\tau_3(\epsilon)\Omega_3^2(\epsilon) d\epsilon,
\end{equation}
where $D(\epsilon)$ is the effective-mass approximation
density-of-states in the GaAs, $f(\epsilon)$ is the Fermi-Dirac
distribution function, $\tau_3$ is the $l = 3$ component in the
multipole expansion of the momentum scattering time, and
$\Omega_3(\epsilon)$ is the $l = 3$ component of the
energy-dependent effective SOI magnetic field. The cubic symmetry
of the Dresselhaus interaction in bulk GaAs \cite{dresselhaus1955}
results in $\Omega_l^2 = 0$ for all $l \neq 3$. Eq.~\ref{taus} is
a generalization of the original DP expression $\tau_s^{-1} = a
\langle \Omega^2 \rangle \tau_p$ \cite{DP1971,MeierZak1984}, where
the integral over energy in Eq.~\ref{taus} properly weights the
spin relaxation rate to account for an arbitrary degree of
degeneracy as well as energy-dependent momentum scattering
mechanisms.

In \textit{n}-GaAs, the dominant scattering mechanism changes from
ionized-impurity (II) scattering at low temperatures to
optical-phonon (OP) scattering at high temperatures
\cite{Fletcher1972a}, as demonstrated by the non-monotonic
temperature-dependence of the mobility shown in Fig.
\ref{fig:chargetransport}(a). To determine the momentum scattering
time, the experimental mobility $\nu$ is fit to the form
\begin{equation}\label{mobility}
\nu^{-1} = {\underbrace{(A+B
T^{3/2})}_{\nu_{\text{II}}}}^{-1}+{\underbrace{(CT^{-1})}_{\nu_{\text{OP}}}}^{-1},
\end{equation}
which combines the II and OP scattering rates via Matthiessen's
rule. In Eq.~\ref{mobility}, A and B are fitting parameters for
the II mechanism and C is a fitting parameter for the OP
mechanism. For II scattering, $T^{3/2}$ is the known temperature
dependence of the scattering time \cite{Brooks1955} and the
fitting parameter $A$ is added to account for degeneracy at low
temperatures. No universal energy exponent can be assigned to OP
scattering over the experimental temperature range, due to the
breakdown of the relaxation-time approximation
\cite{Howarth1953,Fletcher1972a}. We find, however, that $\nu
\propto T^{-1}$ approximates the measured high temperature
mobility. This is not a rigorous relation for OP scattering, but
the purpose of Eq.~\ref{mobility} is to provide a phenomenological
scattering rate which \textit{decreases} with temperature (II
scattering) and a scattering rate which \textit{increases} with
temperature (OP scattering). The fit to Eq.~\ref{mobility} is
shown along with the measured mobility in Fig.
\ref{fig:chargetransport}(a).

After fitting the temperature dependence of the mobility to extract the contributions due to the II
and OP scattering mechanisms, each mechanism is separately fit to the expression
\begin{equation}\label{tau3}
\nu_{\text{II(OP)}}  = \frac{e}{m^* n}\int
D(\epsilon)f(\epsilon)[1-f(\epsilon)]\tau_{1,\text{II(OP)}}(\epsilon)\frac{\epsilon}{k_b
T}d\epsilon
\end{equation}
to determine $\tau_1$ (the momentum relaxation time) for each
mechanism, at each temperature. The energy dependence of the
scattering time is assumed to be $\tau_1 = a \epsilon^\gamma$,
where $\gamma = 3/2$ and $\gamma = 1/2$ for II and OP scattering,
respectively \cite{Lau2004}. The relevant multipole component of
the scattering time for DP relaxation, $\tau_3$, can be determined
from $\tau_1$ by expressing the $l^{th}$ multipole component of
the scattering time using the known form of the scattering cross
section $\sigma(\theta, \epsilon)$
\begin{equation}\label{taul}
\tau_l^{-1}(\epsilon) = \int_{0}^\pi\sigma(\theta,
\epsilon)[1-P_l(\text{cos} \theta)]\sin\theta d\theta,
\end{equation}
where $P_l$ is the Legendre polynomial of degree $l$.
Eq.~\ref{taul} may be evaluated to relate $\tau_3$ to $\tau_1$
(for detailed evaluation of Eq.~\ref{taul} see
Ref.~\onlinecite{MeierZak1984}, resulting in $\tau_1 = \tau_3/6$
for II scattering, and $\tau_1 = 6\tau_3/41$ for OP scattering
\cite{MeierZak1984, Lau2004}).

After fitting the measured mobility with Eq.~\ref{mobility} and
\ref{tau3}, the $l = 3$ component of the momentum scattering rate
$\tau_3^{-1} = \tau_{3,\text{II}}^{-1} +\tau_{3,\text{OP}}^{-1}$
is input to Eq.~\ref{taus}, and the DP spin relaxation rate is
evaluated at all temperatures. The SOI strength used to evaluate
$\Omega_3^2$ as a function of carrier energy is taken from the
$k\cdot p$ calculation with a full fourteen band basis done by
\citet{Lau2001}. Their calculations give $\mathbf{\Omega} =
2\beta/\hbar
(\mathbf{k_x}(k_y^2-k_z^2)+\mathbf{k_y}(k_z^2-k_x^2)+\mathbf{k_z}(k_x^2-k_y^2))$
with $\beta = $ 25 eV \AA$^3$.  The final result for the spin
lifetime as a function of temperature from Eq.~\ref{taus} is shown
as the blue solid line in Fig. \ref{tauetaalpha}(a).
\section{Discussion}

As shown in Fig. \ref{fig:sepdep}, the spin diffusion length
$\lambda = \sqrt{D \tau_s}$ falls from approximately 7 $\mu$m at
20 K to 1 $\mu$m at room temperature. Injector-detector
separations less than approximately 1.0 $\mu$m are therefore ideal
to detect NLSV signals in \textit{n}-GaAs at room temperature. We
emphasize that a two-dimensional model of spin diffusion is needed
to fit the separation dependence of $\Delta V_{nl}$ when the spin
diffusion length is smaller than the channel depth of 2.5 $\mu$m.
Fits using the 1D solution of Eq.~\ref{DD} underestimate the spin
lifetime and spin diffusion length when the channel thickness is
greater than a spin diffusion length, because the spin
accumulation in two dimensions decays faster than $e^{-y/\lambda}$
away from the injector.

As can be seen in Fig. \ref{tauetaalpha}(a), the temperature
dependence of the spin lifetime agrees well with the DP
prediction, calculated from Eq.~\ref{taus}, over the entire
temperature range. $\tau_s$ varies from 49$~\pm~$16~ns at 20~K to
86$~\pm~$10~ps at 300~K. The relatively large uncertainty in the
20~K spin lifetime value results from a lack of data for
injector-detector separations larger than the spin diffusion
length at low temperature. Separations larger than 10~$\mu$m would
be required to constrain the fit adequately. At low temperatures
(40-120~K) we have also measured $\tau_s$ by the NLH measurement.
The spin lifetimes obtained with NLH measurements are also shown
on Fig. \ref{tauetaalpha}(a), with the NLH field sweeps and fits
to Eq.~\ref{NLH} shown in the inset. The $\tau_s$ values from NLH
measurements are in good agreement with the IDSD $\tau_s$ values
above $\sim$60 K. At the lowest temperatures (20-40 K), the pulsed
NLH measurement technique may not be sufficient to completely
remove the effects of DNP. A combined model of the
electron-nuclear spin system is needed to adequately model the NLH
measurement in the regime where DNP is significant, as is done in
Refs.~\cite{Salis2009, Chan2009, Harmon2015}.

We now comment on the magnitude of $\Delta V_{nl}$ in the
biased-detector SV measurement. Combining Eq.~\ref{Vnl} and
Eq.~\ref{compress} allows one to determine the spin accumulation
$n_{\uparrow}-n_{\downarrow}$ given $\Delta V_{nl}$, the SV signal
size. The only unknown is $\eta$, the detection efficiency.  In
our devices, we have demonstrated that $\eta$ is a strong function
of detector bias, which complicates the interpretation. Because of
the detector bias dependence of $\eta$ implied by the data shown
in Fig. \ref{fig:BDSV}, we also cannot assume $\alpha = \eta$, as
the injector contact is biased with a large current, while the
detector bias is varied. Based on these considerations, the spin
polarization of the channel and the injection efficiency may only
be quantitatively evaluated up to a factor of $\eta$
(\textit{i.e.} $\eta P_0$ and $\eta\alpha$, respectively), where
$\eta$ is the detection efficiency at the detector bias voltage at
which the measurement was performed and $P_0$ is the spin
polarization below the injector. These quantities are shown in
Fig. \ref{tauetaalpha}(b). Although the overall scale for $\eta$
cannot be determined in this experiment, it is believed to be
$\sim$50\% based on spin-LED measurements on similar Fe/GaAs
Schottky interfaces \cite{Adelmann2005}.

At the lowest temperatures, we measure $\Delta V_{nl}$ values of
$\sim$1~mV with a forward bias applied to a detector contact. This
implies that the spin-resolved electrochemical potential splitting
at the injector is comparable to the Fermi energy in the GaAs
channel, which is $\sim$5 meV with respect to the conduction band
minimum. As the maximum possible value of $\eta$ is unity, we
emphasize that the ordinate scales shown in Fig.
\ref{tauetaalpha}(b) are therefore minimum values for $P_0$ and
$\alpha$. At 20 K, we measure $\eta P_0 = 30$\%. Thus, the upper
limit of 100\% polarization in the GaAs puts a \textit{lower}
limit of $\eta \sim$ 0.3 at 20 K. Notably, because the forward
bias current (spin extraction) leads to drift \textit{enhancement}
of the spin accumulation buildup at the injector contact, ideal
ferromagnetic contacts ($\alpha =$~1) are not necessary to achieve
channel spin polarizations approaching 100\% \cite{Petukhov2007,
Yu2002}.

In Fig. \ref{tauetaalpha}(b), a downturn in the
injection-detection efficiency product $\eta \alpha$ is observed
at temperatures below 100 K. To address this observation, we have
measured $\eta\alpha$ for different injector current biases. The
results of this measurement are shown in Fig.
\ref{tauetaalpha}(b), where it is apparent that $\eta\alpha$ is a
function of the injector current bias at low temperatures. At
temperatures above $\sim$150 K, where the spin accumulation is
small with respect to the carrier density, $\eta\alpha$ becomes
independent of injector current bias.

To understand the injector bias current dependence of
$\eta\alpha$, we first discuss the influence of an electric field
on the spin accumulation. Electric fields at the injector
necessarily accompany the bias current. In addition to the drift
effects, discussed above, large electric fields in \textit{n}-GaAs
are known to enhance the spin relaxation rate. In \textit{n}-GaAs,
at low temperatures (T $\lesssim$ 30 K) the itinerant electron
temperature can deviate significantly from the lattice temperature
due to the dominance of elastic scattering mechanisms, which
hinder electron-lattice equilibration \cite{Oliver1962}. This
electron heating is present above electric fields $\sim$10 V/cm,
and leads to donor impact ionization, which prevents the electron
temperature from cooling below the donor binding energy ($\sim$6
meV for Si in GaAs \cite{Yu1996}). At low temperatures, electric
field dependence of the spin lifetime has been widely reported
\cite{Kato2004a,Beck2006,Furis2006}. At the lowest temperatures in
our experiment (20, 30 K), the suppression of the spin lifetime
due to the applied electric field may contribute  to the downturn
in $\eta\alpha$ we observe. However, the injector bias dependence
of $\eta\alpha$ is observed clearly up to $\sim$100 K in Fig.
\ref{tauetaalpha}(b). At 100 K, all donors are thermally ionized
and inelastic electron-phonon relaxation mechanisms are sufficient
to prevent any electron-lattice temperature difference. Thus, we
believe that electric field suppression of the spin lifetime is
not the origin of the injector bias dependence of $\eta\alpha$.

We believe that the downturn in $\eta \alpha$ at low temperatures
is more likely to be a consequence of the large spin polarization
of the channel and consequent breakdown of the ordinary
drift-diffusion model. In the presence of a spin accumulation
comparable to the carrier density, Eq.~\ref{DD} must be modified
to prevent the spin polarization from achieving non-physical
values $>$ 100\%. Physically, the model parameters themselves
become functions of the spin polarization, and the assumption of
linear response breaks down \cite{Qi2006}. To be specific, it
becomes necessary to specify the diffusion constants and spin
relaxation rates separately for minority and majority spin
carriers, \textit{i.e.} $\tau_{\uparrow\downarrow}^{-1} \neq
\tau_{\downarrow\uparrow}^{-1} \neq \tau_{s,0}^{-1}/2$ and
$D_{\uparrow} \neq D_{\downarrow} \neq D_0$, where
$\tau_{s,0}^{-1}$ and $D_0$ are the equilibrium spin relaxation
rate and diffusion constant, respectively \cite{Vera-Marun2012}.
We note that for the DP spin relaxation mechanism ($\tau_s^{-1}
\sim \epsilon^3 \tau_p$) in \textit{n}-GaAs where II scattering is
dominant ($\tau_p \sim \epsilon^{3/2}$) the spin relaxation rate
is a strong function of carrier energy $\epsilon$. The diffusion
constant also increases with increasing carrier energy via the
Einstein relation (Eq.~\ref{Einstein}). The mechanisms described
above may provide feedback to limit the spin polarization in the
large spin polarization regime via more efficient spin diffusion
and spin relaxation processes compared to the small spin
polarization linear-response limit.  If this were the case, then
the injector current polarization required to achieve a given spin
accumulation would be larger than that calculated under the
assumption of linear response.

\section{Conclusions}

In conclusion, we have explored several aspects of spin transport
in epitaxial FM/\textit{n}-GaAs spin valves over a wide range of
temperature and bias conditions.  Because these devices are based
on Schottky tunnel barriers, both the injection and detection
efficiencies depend on the bias.  We have exploited this property
to enhance the sensitivity to spin accumulation by applying a bias
current to the detector in the nonlocal configuration.  Although
the mechanism for the enhancement is not well-understood (except
for the role of drift), this approach enables detection of spin
accumulation up to room temperature. At injector current densities
of $10^3$~A/cm$^2$ nonlocal voltages of order $\sim$1~mV are
detected at low temperature, which fall to $\sim$40 $\mu$V at room
temperature. This approach has enabled measurements of the spin
relaxation rate and diffusion length over the entire temperature
range, and good agreement is obtained with a model based  on the
Dyakonov-Perel spin relaxation mechanism. At the lowest
temperatures, however, the standard drift-diffusion model appears
to break down because of the large spin accumulation, which is
comparable to the carrier density. At high temperatures, the
devices are limited by the rapidly increasing spin relaxation
rate, although the injected current polarization also decreases by
a factor of three between 20 K and room temperature.

The devices discussed in this paper are based on Heusler alloys,
which are predicted to have a high spin polarization and grow
epitaxially on GaAs (001).  There is sufficient uncertainty in the
derived values of the detection efficiency and injected current
polarization that it is not possible to make a statement about the
polarization of the Co$_2$FeSi injector beyond the lower bound
(30\%) set by the size of the nonlocal voltage at the lowest
temperature. As suggested by the bias dependence, there is likely
a significant contribution to the tunnelling current from
interface states, a property that is shared by the epitaxial
Fe/GaAs system \cite{Chantis2007}.  Although these important
details still need to be resolved, this work demonstrates that
epitaxial FM/III-V heterostructures can be used to probe spin
transport at room temperature.

\section{Acknowledgments}

This work was supported by the National Science Foundation (NSF)
under DMR-1104951, C-SPIN, one of the six centers of STARnet, a
SRC program sponsored by MARCO and DARPA, the Materials Research
Science and Engineering Centers (MRSEC) program of the NSF under
DMR 08-19885, and the NSF NNCI program.

\bibliography{room_temp_paper}
\end{document}